\documentclass[journal,comsoc]{IEEEtran}
\usepackage{amsmath}
\usepackage{amsfonts}
\usepackage{amsthm}
\usepackage{amssymb}
\usepackage{bbm}
\usepackage[table]{xcolor}
 \usepackage{graphicx}
\usepackage{algorithmic}
\usepackage{algorithm}
\usepackage{xcolor}
\usepackage{multirow}
\usepackage{subfig}
\usepackage{stfloats}

\usepackage{array}
\usepackage{textcomp}
\usepackage{stfloats}
\usepackage{url}
\usepackage{graphicx}
\usepackage{cite}

\usepackage{multirow}
\usepackage{pgfplots}
\usepackage{filecontents}

\usepackage{soul}

\usepackage{hyperref}
\usepackage[numbers,sort&compress]{natbib}

\DeclareGraphicsExtensions{.png,.eps,.ps,.pdf}

\usepackage{tikz}
\usetikzlibrary{arrows,    % Arrow types
                shapes, % Block shapes
                positioning, % Node positioning
                fit,    % Node fit
                backgrounds, % backgrounf
                mindmap,
                petri,
                intersections, % Intersections
                external % 
                }
\definecolor{emerald}{RGB}{69,155,61}
\definecolor{gold}{RGB}{244,216,51}
\definecolor{pink}{RGB}{235,44,206}
\usepackage{verbatim}

\tikzstyle{int}=[draw, fill=cyan!20, minimum size=2em]
\tikzstyle{int_blue}=[draw, fill=blue!20, minimum size=2em]
\tikzstyle{int_green}=[draw, fill=green!20, minimum size=2em]
\tikzstyle{int_red}=[draw, fill=red!20, minimum size=2em]
\tikzstyle{init} = [pin edge={to-,thin,black}]

\usepackage{svg}
\usepackage{filemod}
\usepackage{pgfplots}

\pgfplotsset{
    box plot/.style={
        /pgfplots/.cd,
        black,
        only marks,
        mark=-,
        mark size=1em,
        /pgfplots/error bars/.cd,
        y dir=plus,
        y explicit,
    },
    box plot box/.style={
        /pgfplots/error bars/draw error bar/.code 2 args={%
            \draw  ##1 -- ++(1em,0pt) |- ##2 -- ++(-1em,0pt) |- ##1 -- cycle;
        },
        /pgfplots/table/.cd,
        y index=2,
        y error expr={\thisrowno{3}-\thisrowno{2}},
        /pgfplots/box plot
    },
    box plot top whisker/.style={
        /pgfplots/error bars/draw error bar/.code 2 args={%
            \pgfkeysgetvalue{/pgfplots/error bars/error mark}%
            {\pgfplotserrorbarsmark}%
            \pgfkeysgetvalue{/pgfplots/error bars/error mark options}%
            {\pgfplotserrorbarsmarkopts}%
            \path ##1 -- ##2;
        },
        /pgfplots/table/.cd,
        y index=4,
        y error expr={\thisrowno{2}-\thisrowno{4}},
        /pgfplots/box plot
    },
    box plot bottom whisker/.style={
        /pgfplots/error bars/draw error bar/.code 2 args={%
            \pgfkeysgetvalue{/pgfplots/error bars/error mark}%
            {\pgfplotserrorbarsmark}%
            \pgfkeysgetvalue{/pgfplots/error bars/error mark options}%
            {\pgfplotserrorbarsmarkopts}%
            \path ##1 -- ##2;
        },
        /pgfplots/table/.cd,
        y index=5,
        y error expr={\thisrowno{3}-\thisrowno{5}},
        /pgfplots/box plot
    },
    box plot median/.style={
        /pgfplots/box plot
    }
}

\begin{document}

%\title{Coverage Increase in 5G mmWave by Dynamic Waveform Switching}
%\title{On the system performance of 5G Uplink Access Schemes}

\title{AI-Assisted Dynamic Port and Waveform Switching for Enhancing UL Coverage in 5G NR}

\author{Alejandro Villena-Rodríguez, Gerardo Gómez, Mari Carmen Aguayo-Torres, Francisco J. Martín-Vega, %<- 
    \\José Outes-Carnero, F. Yak Ng-Molina, Juan Ramiro-Moreno%<-
\thanks{Manuscript received June xx, 2024; revised XXX. %
This work has been funded by Ericsson, the European Fund for Regional Development (FEDER), and AEI (Spain), through grants RYC2021-034620-I, and by MCIN/AEI/10.13039/501100011033 through grant PID2020-118139RB-I00.}
\thanks{Villena-Rodríguez, Gómez, Aguayo-Torres and Martín-Vega are with the Communications and Signal Processing Lab, Telecommunication Research Institute (TELMA), Universidad de M\'alaga, E.T.S. Ingenier\'ia de Telecomunicaci\'on, Bulevar Louis Pasteur 35, 29010 M\'alaga (Spain). 
Outes-Carnero, Ng-Molina, and Ramiro-Moreno are with Cloud and Software Services at Ericsson, 29590, Malaga, Spain. (e-mail: avr@ic.uma.es)}
%\thanks{This work has been submitted for publication. Copyright may be transferred without notice, after which this version may no longer be accessible.}
}

% Conference style:
% \author{
% \authorblockN{Alejandro Villena-Rodríguez$^{(1)}$, Gerardo Gómez$^{(1)}$, Mari Carmen Aguayo-Torres$^{(1)}$, Francisco J. Martín-Vega$^{(1)}$}
% \authorblockN{José Outes-Carnero$^{(2)}$, Adriano Mendo $^{(2)}$, Juan Ramiro-Moreno$^{(2)}$}
% \authorblockA{\{avr, ggomez, aguayo, fjmvega\}@ic.uma.es, \{jose.outes, adriano.mendo, juan.ramiro\}@ericsson.com}
% \authorblockA{$^{(1)}$Instituto Universitario de Investigaci\'on en Telecomunicaci\'on (TELMA), Universidad de M\'alaga, Spain.}
% \authorblockA{$^{(2)}$ Ericsson, Málaga, Spain.}
% }

%\thanks{The authors are with the Communications and Signal Processing Lab, Instituto Universitario de Investigaci\'on en Telecomunicaci\'on (TELMA), Universidad de M\'alaga, CEI Andaluc\'ia TECH, ETSI Telecomunicaci\'on, Bulevar Louis Pasteur 35, 29010 M\'alaga (Spain) (e-mail: \{mls, avr, ggomez, fjmvega, aguayo\}@ic.uma.es)}% <-this % stops a space

\maketitle

\begin{abstract} % [Fran]: Para IEEE Wireless Communications letters (Q1) & IEEE Communications Letters (Q2) el abstract son de 75 a 100 palabras. 
% [Fran] Voto por llamar throughput al throughput, ya que mucha gente no conoce ese término, y en el 3GPP, en los ejemplos de 5G de MATLAB, etc., llaman throughput a los bits/sec correctamente decodificados.
The uplink of 5G networks allows selecting the transmit waveform between cyclic prefix orthogonal frequency division multiplexing (CP-OFDM) and discrete Fourier transform spread OFDM (DFT-S-OFDM), which is appealing for cell-edge users using high-frequency bands, since it shows a smaller peak-to-average power ratio, and allows a higher transmit power. Nevertheless, DFT-S-OFDM exhibits a higher block error rate (BLER) which complicates an optimal waveform selection. In this paper, we propose an intelligent waveform-switching mechanism based on deep reinforcement learning (DRL). In this proposal, a learning agent aims at maximizing a function built using available throughput percentiles in real networks. Said percentiles are weighted so as to improve the cell-edge users' service without dramatically reducing the cell average. Aggregated measurements of signal-to-noise ratio (SNR) and timing advance (TA), available in real networks, are used in the procedure. Results show that our proposed scheme greatly outperforms both metrics compared to classical approaches.
\end{abstract}


\section{Introduction}
\IEEEPARstart{T}{o} overcome the limited transmission power and power backoff of the user equipment's (UEs) power amplifiers (PAs), the
% One of the key pillars of 5G New Radio (NR) communications is the transmission over millimeter Wave (mmWave) bands. While this part of the spectrum increases the available bandwidth, the maximum cell range tends to be shorter due to a higher path loss \cite{ref:Giordani19}. This path loss is more challenging in the uplink (UL) than in the downlink (DL) due to: (i) the maximum transmit power of the user equipment (UE) is smaller than that of the base station (BS), and (ii) the input power backoff of power amplifiers (PAs) is smaller in the UE than in the BS. These issues derive from the fact that the UE is constrained to be portable equipment with a reduced cost. %However, the complexity of the receivers can be higher in the UL due to BSs hardware having fewer energy constraints and greater computational power.
%
3GPP specifications introduced two possible waveforms in the uplink (UL) of 5G: cyclic prefix orthogonal frequency division multiplexing (CP-OFDM) and discrete Fourier transform spread OFDM (DFT-S-OFDM) \cite{ref:MPR}. %, which is also known as \textit{transform precoding} in the 3GPP standard. 
The former leads to higher spectral efficiency and better performance of multiple input multiple output (MIMO) techniques whereas the latter
leads to a smaller peak-to-average power ratio (PAPR). As shown in preliminary studies \cite{ref:Zheng09}, in the presence of non-linear PAs and UL power control, DFT-S-OFDM leads to fewer errors and higher throughput compared to CP-OFDM for UEs placed at distant locations from its serving base station (BS), whereas CP-OFDM offers a higher throughput for the rest of UEs. The fact that out-of-band emissions are smaller with DFT-S-OFDM implies that it is possible to transmit with a higher power than with CP-OFDM while fulfilling the same transmission spectral mask. 
Hence, the 3GPP NR specifications define a maximum transmit power of DFT-S-OFDM, which is between $1.5$ and $2.5$ dB greater than that of CP-OFDM, depending on the bandwidth and constellation \cite{ref:MPR}. This higher transmit power extends the cell coverage when DFT-S-OFDM is used, which can potentially increase the low percentiles of the throughput distribution in a cell. 

% [Fran] Cambio el texto de abajo para queue aparezca redactado
% 1 )DFT-S-OFDM also shows a smaller peak-to-average power ratio (PAPR) as compared to CP-OFDM. As shown in preliminary studies \cite{ref:Zheng09}, in the presence of non-linear PAs and UL power control, DFT-S-OFDM leads to fewer errors and higher throughput compared to CP-OFDM for UEs located distant from its serving BS. 

% 1a) Out-of-band emissions are smaller with the latter, making it possible for DFT-S-OFDM to transmit with higher power while still fulfilling the same transmission spectral mask. 

% 2) The 3GPP NR specifications describe the available extra power as a value between $1.5$ and $2.5$ dB depending on the bandwidth and constellation \cite{ref:MPR}. 

% 3) In addition, the extra transmitting power improves signal reception, thus extending the cell coverage. In general, DFT-S-OFDM is more commonly used by those users in the lower percentiles of the throughput distribution in a cell. 

% ANTIGUO: An interesting invention in this topic is described in \cite{ref:WO2021/047973}, where two mechanisms for waveform switching are proposed to optimize the power consumption of the UE. One is based on a change in the Radio Resource Control (RRC) configuration through the delivery of an RRC reconfiguration message that changes the UL waveform; and the other is to perform a layer 1 (L1) bandwidth part (BWP) switching. 

This potential benefit is exploited in  \cite{ref:WO2021/047973}, where a mechanism for waveform switching is proposed to optimize the power consumption of the UE. This invention, named Dynamic Port and Waveform Switching (DPWS), relies on the transmission of Radio Resource Control (RRC) configuration messages to switch the UL waveform. 
 Nevertheless, this switching mechanism has a cost in terms of time, since the communication is interrupted during a guard time after every switch. Therefore a mechanism based on counters and timers is proposed to avoid a ping-pong effect for UEs whose performance is similar with both waveforms. By intelligently selecting the UL waveform, the proposed solution can potentially boost the UEs' power efficiency while improving throughput and error rate. However, it also poses the challenge of selecting the switching parameters, which is still an open research problem. % [Fran] Justo arriba decimos que este waveform switching mechanism puede reducir la potencia transmitida. Esto puede encenderle la bombilla a los revisores y que nos pidan que demostremos que el UE ahorra potencia...igual es mejor no mencionar eso y hablar solo del throughput. [MC] Pero si al pasar al DFT-S-OFDM enviamos 2dBs más, no creo que se use menos potencia en realidad, ¿no? Si acaso será más eficiente su uso. A lo mejor es como se puede poner

In this context, deep reinforcement learning (DRL) appears as a promising branch of artificial intelligence (AI) that has the potential to overcome such complex interplays between different metrics, offering real-time inference capabilities \cite{ref:Feriani21}. %compared to traditional approaches based on convex optimization \cite{ref:Feriani21}. 
DRL frameworks allow for an optimization of an arbitrary objective, namely a reward function, by using a fitting policy for a particular problem. In the context of this paper, the suitableness of DRL comes from the fact that said policy can be learned while interacting with the system by applying changes and observing its effects. %, rather than relying on access to ground-truth data. 
Thanks to these benefits DRL techniques have been recently applied to optimize the UL of 5G networks, e.g., in \cite{Ahsan21} DRL is used to determine an optimal user clustering for non-orthogonal access that maximizes the throughput, whereas in \cite{Neto21} and \cite{Teng19}  DRL is used to determine the optimal UL control commands to maximize the throughput and reduce delay respectively.

Despite of their relevance, previous works related to UL optimization do not treat the problem of waveform switching. The only exceptions are \cite{ref:MPR}\cite{ref:Zheng09}, but they fail to provide a mechanism to select the switching configuration autonomously. %, without human intervention. 
To the authors' knowledge, there is no proposal in the literature presenting an autonomous waveform switching optimization mechanism for the UL of 5G and 6G systems. 

In this paper, we present a novel solution, named \textit{artificial intelligence-assisted dynamic port and waveform switching} (AI-DPWS), whose objective is to find the optimal configuration parameters that govern the switching of the UE waveform and transmission antenna ports. The proposed AI-based framework finds optimal values for the signal-to-noise ratio (SNR) threshold and the appropriate SNR hysteresis that maximize the cell's performance in terms of UL throughput. 

More precisely, AI-DPWS aims to improve the throughput of the cell-edge UEs, i.e., those associated with a lower percentile of the cell´s throughput cumulative distribution function (CDF), without sacrificing the throughput of the cell-interior UEs (higher percentiles). This is achieved through key performance indicators (KPIs) that are available in real-world networks, such as histograms of the throughput, UL SNR, and timing advance (TA), which are collected during real-life network operation. 
The aim of considering these realistic metrics is to offer a solution that can be deployed in real cells. 

%AQUI FALTA QUÉ OPTIMIZAMOS, TANTO QUE QUEREMOS MEJORAR EL THROUGHPUT COMO QUE LO HACEMOS CON MEDIDAS REALISTAS, PRESENTES EN LAs REDES ACTUALES. ASÍ, USAMOS EL TIMING ADVANCED, NO LA DISTANCIA, QUE NO ES MEDIBLE, Y LA SNR. ADEMÁS, EMPLEAMOS MEDIDAS AGREGADAS, NO POR USUARIO, QUE NO ESTÁN DISPONIBLES EN UNA RED REAL. ESTO ES IMPORTANTE PORQUE ES MUCHO MÁS REALISTA. %Interestingly, this improvement affects not only the average throughput but also the 10-th percentile of the throughput, which represents the performance of the worst users. Thus, the fairness of the system is improved thanks to this mechanism. % and UE power consumption. % [Fran] Nos pueden pedir que demostremos que se ahorra power consumption si ponemos eso. Tenemos que esta seguros de que se ahorre efectivamente potencia con el swicthing. 

% The rest of the paper is organized as follows. Section \ref{System_model} summarizes the system model. The considered DPWS scheme is described in section \ref{DPWS}. The proposed DRL framework for DPWS, named AI-DPWS, is described in section \ref{DRL_framework}. Simulation results are discussed in Section \ref{Simulationresults}. Finally, section \ref{conclusions} draws the main conclusions of this work.

\section{System Model} 
\label{System_model}
For each UL transmission, we investigate a link between a single cell and a single user in the UL direction, i.e. without interfering UL transmission, considering the 3GPP specifications for the physical and medium access control (MAC) layers of 5G. 
Therefore the transmission chain uses the low-density parity-check codes (LDPC) for the physical uplink shared channel (PUSCH). 
The proposed system allows selecting between two different UL waveforms: i) DFT-S-OFDM waveform with 1 antenna port and 1 data layer, and \mbox{ii) CP-OFDM} waveform with 2 antenna ports and 1 data layer. It is worth noting that the DFT-S-OFDM scheme only makes use of 1 antenna port. This design decision was taken due to the moderate to low compatibility of DFT-S-OFDM with multiple-input multiple-output (MIMO) \cite{ref:Zaidi}. Both modulation schemes use demodulation reference signals (DMRS) to perform channel estimation for symbol detection (i.e., equalization). DMRS signals might be precoded by a transmit precoding matrix in the case of CP-OFDM. Nevertheless, we also consider the periodic transmission of sounding reference signals (SRS), which are not precoded, to estimate the SNR for waveform selection. 
% [Fran] 2)	Nos van a preguntar que por qué NO usamos 2 puertos con DFT spread. ¿Es fácil lanzar simulaciones con DFT spread de 2 puestos y con OFDM de un solo puerto? 

\begin{figure}[t]
\centering
\includegraphics[width=1.0\columnwidth]{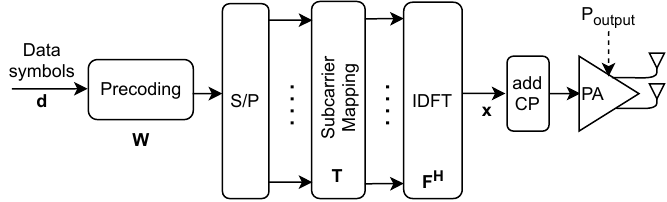}
\caption{Digital base-band transmitter structure of CP-OFDM.}
\label{cpofdm_transmitter}
\end{figure}

\begin{figure}[t]
\centering
\includegraphics[width=1.0\columnwidth]{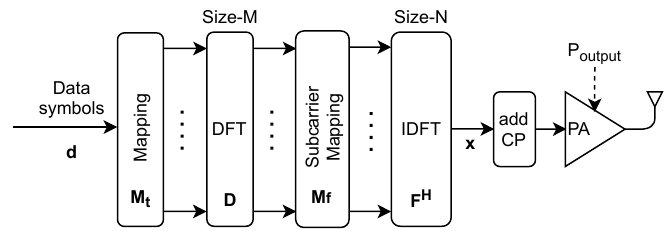}
\caption{Digital base-band transmitter structure of DFT-S-OFDM.}
\label{dftsofdm_transmitter}
\end{figure}

Fig. \ref{cpofdm_transmitter} and Fig. \ref{dftsofdm_transmitter} show the digital base-band transmitter structure of CP-OFDM and DFT-S-OFDM, respectively, being the elements of vector $\mathbf{d} \in \mathbb{C}^{1 \times N_{d}}$ a set of complex data symbols to be transmitted, with $N_d$ the number of symbols.

% For simplicity, the precoding matrix considered in this paper is built for 1 data layer and 2 antenna ports as $\mathbf{W} = [1 \quad 1]^T / \sqrt{2}$. 

As shown in Fig. \ref{cpofdm_transmitter}, for the CP-OFDM signal generation, data symbols are transformed via a precoding matrix $\mathbf{W} \in \mathbb{C}^{N_{\rm tx}\times 1}$ where $N_{\rm tx}$ is the number of antenna ports. This precoding matrix maps data layers onto the number of antenna ports. After the multiplication with the matrix $\mathbf{W}$, the resulting data symbols are transposed. Then, the precoded and transposed symbols are passed through a subcarrier mapping matrix $\mathbf{T} \in \mathbb{C}^{N\times N_d}$, being $N$ the number of subcarriers. Finally, the output of the matrix $\mathbf{T}$ is converted to the time domain via $\mathbf{F}^H$ where $\mathbf{F}^H \in \mathbb{C}^{N\times N}$ is an inverse discrete Fourier transform (IDFT). The final signal $\mathbf{x} \in \mathbb{C}^{N\times N_{\rm tx}}$ in the time domain is generated as following: $\mathbf{x} = \mathbf{F^H T (W d)^T}$. Afterward, a cyclic prefix is added to the resulting time signal before being fed into the non-linear PA. 

% wliufghuwig

Regarding the DFT-S-OFDM signal generation (Fig. \ref{dftsofdm_transmitter}), data symbols are mapped onto a DFT matrix denoted by $\mathbf{D} \in \mathbb{C}^{M\times M}$ via a mapping matrix $\mathbf{M_t} \in \mathbb{C}^{M\times N_d}$, where $M$ is the DFT size. Then, the output of the DFT is mapped onto a set of subcarriers in the frequency domain through another mapping matrix $\mathbf{M_f} \in \mathbb{C}^{N\times M}$. Finally, the output of the matrix $\mathbf{M_f}$ is converted to time domain via $\mathbf{F}^H$, where $\mathbf{F}^H \in \mathbb{C}^{N\times N}$ is the IDFT matrix and $N$ is the number of subcarriers. The final signal $\mathbf{x}\in \mathbb{C}^{N\times 1}$ in the time domain is generated as follows: $\mathbf{x} = \mathbf{F}^H \mathbf{M_f D _t M_t d}$. Then, a CP is added to the resulting time signal before being fed into the non-linear PA. 

In both cases, output signals are amplified by a non-linear PA, which is assumed to be memory-less with amplitude-to-amplitude distortion only. In particular, a Rapp model is considered \cite{ref:RappAmp}, whose amplitude-to-amplitude conversion function is given by: 
\begin{equation}\label{eq:rapp}
g(v,A) = vA\left( {1 + {\rm{abs}}\left( {\frac{{vA}}{{A_{{\rm{sat}}} }}} \right)^{2p} } \right)^{\frac{{ - 1}}{{2p}}}, 
\end{equation}

\noindent
where $v$ is the gain of the small-signal, $A$ is the amplitude of the input signal, $A_{\rm sat}$ is the limiting output amplitude, and $p$ controls the smoothness of the transition from the linear region to the \textcolor{black}{saturation regime}. 

A closed-loop power control mechanism is used in the UL \cite{ref:38.331}, %of the Physical Uplink Shared Channel (PUSCH) \cite{ref:Simonsson}
which considers that the BS decides a transmit power, $P_{\rm tx}$, to compensate for the path loss, $\mathrm{PL}$. %. This decided transmit power is based on the sum of two parts\textcolor{black}{:} An open power control part, that the UE computes based on DL reference signals; and an offset that is based on Transmit Power Control (TPC) commands received from the BS on the UL grant assignments. After a number of TPC receptions, the decided transmit power converges to the following value:
This transmit power can be expressed in decibels as $P_{\rm tx} = P_0 + \alpha \mathrm{PL}$ dB, being $P_0$ a target received power at the BS, and $\alpha \in [0, 1]$ the fractional compensation factor. An Urban Macro (UMa) model has been considered for computing the path loss component, PL.
%If $\alpha = 0$, the path loss is not compensated, and thus $P_{\rm decided} = P_0$, whereas with $\alpha = 1$, $P_{\rm decided}$ fully compensates the path loss. 
% QUITAR SI FALTA ESPACIO O SE VE QUE NO APORTA NADA. 
% In principle, $\alpha = 1$ seems beneficial since it fully compensates the path loss; however, distant users also transmit with high power, leading to a strong interference on neighbouring cells. 
%As shown in \cite{ref:Martin16}, values close to $\alpha = 0.75$ maximizes average capacity when the BSs are placed randomly according to a Poisson point process.

Restrictions on the maximum transmit power supported by the UE and also out-of-band emissions, impose further limits on the transmitted power. 
The maximum power reduction (MPR) \cite{ref:MPR} specifies the decrease in the maximum power transmitted in order to enable the device to fulfill the requirements of the transmitter adjacent channel leakage ratio. This value imposes a maximum transmit power to guarantee that the out-of-band emission is below a given threshold. Since these out-of-band emissions depend on the waveform, modulation level and channel bandwidth, the possible MPR values also depend on such parameters. Finally, the final power transmitted by the UE, $P_{\mathrm{output}}$, can be defined as $P_{\rm output} = \min(P_{\rm decided}, P_{\rm max}^{'})$, where $P_{\rm max}^{'} = P_{\rm max} - \mathrm{MPR}$. 

% *************************************************************
% Quito tabla de MPR y discusión por falta de espacio. Si nos dan major podemos meter una página más y metemos esto
% *************************************************************
% Table \ref{MPRtable} summarizes the power reduction values as per the 5G specs. Notice that a higher maximum power can be used with DFT-S-OFDM, since its MPR is smaller than with CP-OFDM. This is an expected result as the former waveform is related to a smaller PAPR. DFT-S-OFDM has shown PAPR values between 7 and 8.5 dB while CP-OFDM produced values between 10 and 11 dB.

% % table: MPR
% \begin{table}[h]
% \renewcommand{\arraystretch}{1.3}
% \caption{MPR values \cite{ref:MPR}}
% \label{MPRtable}
% \begin{center}
% \begin{tabular}{|c|c|c|c|}
% \hline
% \multicolumn {2}{|c}{}& \multicolumn {2}{|c|}{MPR (dB)} \\
% \hline
% Waveform & Modulation & \begin{tabular}[x]{@{}c@{}}50/100/200 MHz\\Channel BW\end{tabular} & \begin{tabular}[x]{@{}c@{}}400 MHz\\Channel BW\end{tabular}\\
% \hline
% \multirow{4}{*}{DFT-S-OFDM} & Pi/2BPSK & 1.5 & 3.0\\\cline{2-4}
% & QPSK & 1.5 & 3.0\\\cline{2-4}
% & 16QAM & 3.0 & 4.5\\\cline{2-4}
% & 64QAM & 5.0 & 6.5\\
% \hline
% \multirow{3}{*}{CP-OFDM} & QPSK & 3.5 & 5.0 \\\cline{2-4}
% & 16QAM & 5.0 & 6.5\\\cline{2-4}
% & 64QAM & 7.5 & 9.0\\
% \hline
% \end{tabular}
% \end{center}
% \end{table}

% *************************************************************
% Table fin
% *************************************************************

The SNR is computed as follows:
% [Fran] echo en falta alguna expresión que determine el h_i en función de H de verdad que tiene el canal y la matriz de precoding...que se de algo más de detalle de qué es lo que representa exactamente. 
\begin{equation}\label{eq:SNReq}
\mathrm{SNR} = P_{\rm output} - \mathrm{PL} + 10\,\text{log}_{10}  \left(\sum_{i=1}^{N_\mathrm{rx}} |h_{i}|^2 \right) - N_0\quad \mathrm{[dB]},
\end{equation}

\noindent where $N_\mathrm{rx}$ is the number of receiver antennas, and $h_{i}$ is the \textit{effective channel} of the $i$-th receiver antenna after precoding (in case of CP-OFDM). %and channel estimation. 
The term $N_0$ represents the thermal noise in the band assigned to the user, given by $N_{0} = -204 + 10\,\text{log}_{10} \left( 12\times \Delta f\times N_{\rm RB} \right) + N_{\rm fig} \quad \mathrm{[dBW]}$, where $ \Delta f$ is the subcarrier spacing \textcolor{black}{in Hz}, $N_{\rm RB}$ is the number of physical resource blocks (PRBs) assigned to the user and $N_{\rm fig}$ is the noise figure at the BS. % {\color{red} [REF]}.

\section{Dynamic Port and Waveform Switching} 
\label{DPWS}

DPWS feature is targeted to switch between CP-OFDM and DFT-S-OFDM to enhance the UL coverage. In order to achieve that, it uses the following parameters: 

\begin{itemize}
    \item \textit{Threshold ($\zeta$)}: it determines switching occasions from CP-OFDM to DFT-S-OFDM when the following inequality is fulfilled $\mathrm{SNR} < \zeta$.
    \item \textit{Hysteresis ($\xi$)}: it determines the switching occasions from DFT-S-OFDM to CP-OFDM as follows $\mathrm{SNR} > \zeta + \xi$.
    \item \textit{Counter ($C$)}: it counts for the number of switching occasions in order to trigger a waveform switching.    %\item 
    \item \textit{Timer ($T$)}: it determines the time window to account for switching occasions. To trigger a waveform switch the number of switching occasions must be equal to $C$ within a time window that is smaller than $T$. This time window is expressed as a number of SRS receptions. 
%\textit{SRS periodicity}: SRS reporting period, which has a direct impact on how often switching occasions can be reported.
\end{itemize}

The switching mechanism also takes into consideration the SNR estimated with the SRS $(\gamma)$ and the current waveform, as follows. If the transmission is done with CP-OFDM, then a switching occasion is counted if the SNR falls below $\zeta$. However, if the transmission is done with DFT-S-OFDM, then a switching occasion is counted if the SNR raises above $\zeta + \xi$. The switching mechanism is detailed in \textbf{Algorithm \ref{alg:DPWSalg}}, \textcolor{black}{which is executed for each SRS reception}. 

\begin{algorithm}

    \textbf{Input:} $\zeta$, $\xi$, \textit{C}, \textit{T}, $\gamma$, CurrentWaveform \\
    \textbf{Initialize:} $c = 0$, $t = 0$
\begin{algorithmic}[1]   
\IF{($t < T$)}
{
   \IF {CurrentWaveform == CP-OFDM and $\gamma <  \zeta$ }
   {
      \STATE $c = c + 1$ 
   }
   \ELSIF { CurrentWaveform == DFT-S-OFDM and $\gamma >  \zeta + \xi$ }
   {
      \STATE $c = c + 1$ 
   }
   \ELSE 
   {
      \STATE $c = 0, \, t = 0$
   }
   \ENDIF

   \IF {$ c \geq {C}$ }
   {
      \STATE Perform waveform and port switching
   }
   \ENDIF
   \STATE $t = t + 1$ 
}
\ELSE
{
    \STATE $c = 0, \, t = 0$
}
\ENDIF
\end{algorithmic}
\caption{DPWS algorithm}
\label{alg:DPWSalg}
\end{algorithm}

% [Fran] Mecanismo de cambio de onda. 
The waveform switch is signaled via an RRC reconfiguration message, that changes the UL waveform from CP-OFDM to DFT-S-OFDM and vice-versa. Since the RRC reconfiguration message can change any parameter of the UL transmission, the 3GPP specifications reserve a guard time that allows the UE to prepare for transmission according to the new configuration. According to  \cite{ref:38.331}% \cite{ref:38.133}
, this guard time depends on the numerology, $\mu$, and on the UE category, but it ranges from $16.75$ ms ($\mu=3$, type 1 category) to $19$ ms ($\mu=0$, type 2 category). Therefore, each switch has a cost in performance since it involves an interruption in the UL transmission, which must be considered by the agent.

\section{Proposed Deep Reinforcement Learning Dynamic Switching} 
\label{DRL_framework}

Our proposed framework, AI-DWPS, is based on a DRL algorithm. In particular, a Deep Q-Learning (DQL) approach is used, where the Q-Table typically used in classical RL algorithms is substituted by a neural network called Q-Network. This Q-Network is implemented by a Multi-Layer Perceptron (MLP) with a number of nodes in the input layer that matches the cardinality of the state space whereas the number of nodes in the output layer matches the cardinality of the action space. It has a single hidden layer with $n_\mathrm{hidden}$ nodes.

%The UE can be asked by the cell to change its UL waveform and port configuration according to the rules set by the DPWS algorithm of the attached cell, which handles the configuration of its own individual set of DPWS parameters. 
In this paper, we propose to optimize the most important DPWS parameters, namely SNR threshold ($\zeta$) and SNR hysteresis ($\xi$). %The rest of the parameters are assumed to be fixed. DPWS algorithm controls when and how the UL waveform and ports of the UE connected to the cell will switch.
These parameters are controlled by a DRL agent, that can decide to modify them. The DPWS configuration of each cell can be changed based on the actions suggested by the DRL agent. The DRL is located in a central network element, e.g., operations support systems (OSS), or in an external server.% (see Fig. \hl{XXX}).

\subsection{States, actions, and reward}

\subsubsection{States}
%In RL algorithms the state is a numerical representation of the situation of the environment that is used by the agent to decide which action it will take. 
Given the nature of the problem, we have chosen to report statistics about two realistic uplink measurements available in actual networks: the SNR and the TA distrubutions. The throughput metric is defined as the rate of correctly decoded bits per second, whereas the TA indicates how distant is the UE from the serving BS.  
Both SNR and TA metrics are not reported on a per-connection (i.e., per-user) basis but as a histogram of all connections in a given time window. 
We have considered $L=12$ bins for TA and SNR. The SNR histogram has the following bins' limits (in dB): $[-\infty, -5, -2, 1, 4, 7, 10, 13, 16, 19, 22, 25, \infty]$; whereas, the bins' limits for the TA histogram are (in $\%$): $[5, 15, 25, 35, 45, 55, 65, 75, 85, 95, 105, 115, \infty]$. Note that the TA bins' limits are expressed in terms of percentages, being $100\%$ the TA associated with the propagation delay at the maximum range of the cell. This cell range is a network parameter representing the maximum distance (in meters) at which a given BS provides coverage. Due to the channel delay spread, the instantaneous measurement of the TA may provide a value above $100\%$ for cell edge users.

%Thus, the bin related to a TA value of $115\%$ represents the number of UEs placed at a distance up to a $15\%$ higher than the maximum configured cell range and/or it´s TA reported value was up to a $15\%$ higher than that of the configured cell range.

To compress the information given by these histograms we define the following descriptors for the TA and SNR metrics:
\begin{equation}\label{eq:relation}
R_{\ell, {\Psi}} =  \frac{\sum_{i = \ell}^{L} \mathrm{BINS}_{i, {\Psi}}}{\sum_{i = 1}^{\ell-1} \mathrm{BINS}_{i, {\Psi}}}, 
\; D_{\ell, {\Psi}} =  \frac{\sum_{i = \ell}^{L} \mathrm{BINS}_{i, {\Psi}}}{\sum_{i = 1}^{\max} \mathrm{BINS}_{i, {\Psi}}},
\end{equation}

% \begin{equation}\label{eq:division}
% D_{\ell, \mathrm{metric}} =  \frac{\sum_{i = \ell}^{\max} \mathrm{BINS}_{i, \mathrm{metric}}}{\sum_{i = 1}^{\max} \mathrm{BINS}_{i, \mathrm{metric}}}
% \end{equation}

\noindent {where $\Psi$ stands for a given metric, i.e., either the SNR or TA, $L$ is the number of bins, and $\mathrm{BINS}_{i, {\Psi}}$ stands for the number of UEs whose metric (TA or SNR) falls within the $i$-th bin interval, which is expressed as $\left[\psi^{-}_i, \psi^{+}_i \right]$}. %divided by the overall number of UEs associated with the serving cell}. 
%These values describe the relationship between low and high values of the SNR and TA. For example, a value of $D_{9,\mathrm{TA}} \simeq 1$ indicates that almost all UEs in the cell are positioned in the cell edge. 

These metrics inform about the SNR and TA histograms while giving information about their shapes. More specifically, $D_{\ell, \mathrm{metric}}$ represents an estimation of the complementary CDF (CCDF), which represents the probability that the random metric, $\Psi$, is higher than the smaller limit of the $\ell$-th bin, i.e., $\hat{\Pr} (\Psi > \psi_\ell^{-})$. Thus, for instance, a value of $D_{9,\mathrm{TA}} \simeq 1$ indicates that almost all UEs in the cell are positioned in the cell edge. 
On the other hand, $R_{\ell, \mathrm{metric}}$, can be understood as the ratio between the number of occurrences of the random metric being above the smaller edge of the $\ell$-th bin, $\psi_\ell^{-}$, and the number of occurrences below such edge of the $\ell$-th bin. Therefore this statistics can be understood as an estimation of this quotient of probabilities, $\hat{\Pr} \left( \Psi \geq \psi_\ell^{-} \right)/\hat{\Pr} \left( \Psi < \psi_\ell^{-} \right)$.

%... ALGO HAY QUE EXPLICAR DE EN QUÉ SON DISTINTAS LAS DOS MEDIDAS Y POR QUÉ NOS INTERESAN POR SEPARADO 

Let $\mathcal{S}$ be the state space, with $s[n] \in \mathcal{S}$ the instantaneous state at step $n$. The instantaneous state for the DLR algorithm is defined as:
% \begin{align}
%     s[n] = \{&\zeta[n], \xi[n],\bar{\gamma}[n], R_{6, \mathrm{SNR}}, \nonumber \\
%          & R_{5, \mathrm{SNR}},R_{6, \mathrm{TA}},R_{3,\mathrm{TA}},D_{5, \mathrm{SNR}}\},
% \end{align}

\begin{align}
    s[n] = \{&\zeta[n], \xi[n],\bar{\gamma}[n], R_{6, \mathrm{SNR}}, R_{5, \mathrm{SNR}},R_{6, \mathrm{TA}},R_{3,\mathrm{TA}},D_{5, \mathrm{SNR}}\},
\end{align}

\noindent being $\bar{\gamma}$ the average SNR \textcolor{black}{across the cell's UEs and slots related to a given step}, $\zeta[n]$ the SNR threshold at step $n$ and $\xi$ the SNR hysteresis at step $n$.% In order to consider good and bad quality users, UN HACER NOTAS  QUE SE COGEN VARIOS DECILES

% Let $G[n]$ be the average user throughput in step $n$, and let $\mathcal{S}$ be the state space, with $s[n] \in \mathcal{S}$ the instantaneous state at step $n$. Then, the  state space is defined as
% \hl{$s[n] = \{G[n], \gamma_0[n], SNRdistr\}$}, being $\gamma_0[n]$ the current SNR threshold and step $n$.

\subsubsection{Actions}

% Los mejores entrenamientos en realidad solo optimizan el threshold. Tengo otros entrenamientos con ambos (thres e histeresis) pero no son tan buenos
Let $\mathcal{A}$ be the action space, with $a[n] \in \mathcal{A}$ being the action chosen by the agent at step $n$. The action defines the decision made for each optimizable parameter: SNR threshold  $(\zeta)$ and SNR hysteresis $(\xi)$. Both parameters are subjected to 3 possible options: i) decrease the value of the parameter by a certain step, ii) keep the existing value, or iii) increase value by the same step. The decrease/increase step is fixed to $\Delta\zeta = 1$ dB and $\Delta\xi = 0.5$ dB. $\mathcal{A}$ is therefore comprised of 9 possible actions. Notice that the use of small changes of $\zeta$ and $\xi$ reduces the impact of wrong decisions made by the agent, and allows to reach smoothly optimal values  through several iterations. %Using higher increments would reduce the number of iterations needed to reach the optimal values at the expense of increasing the risk of overshooting.

\subsubsection{Reward} % [FRAN] SEGUIR!!
To accurately capture the performance of the users in the cell, the reward function is composed of a set of $K$ reward factors, $G_k, k\in[1,K] \subset \mathbb{N}$, each one corresponding to a certain percentile, $p_k$, of the throughput of the cell. 

The reward function at step $n$ is given by
\begin{align}
\label{eq:reward_n}
r[n] &= \theta  \cdot \mathbf{B} \cdot   \left( \Delta G_{1},\Delta G_{2}, ..., \Delta G_{k} \right)^t, \\
\label{eq:gain_percentile}
\Delta G_{k} &= \frac{G_{k}[n] - G_{k}[n-1]}{G_{k}[n-1]},
\end{align}

% \begin{equation}
% \label{eq:reward_n}
% r[n] = \theta  \cdot \mathbf{B} \cdot   \left( \Delta G_{1},\Delta G_{2}, ..., \Delta G_{k} \right)^t, 
% \end{equation}
% %
% \label{eq:gain_percentile}
% \begin{equation}
% \Delta G_{k} = \frac{G_{k}[n] - G_{k}[n-1]}{G_{k}[n-1]},
% \end{equation}

% ANTIGUO
% \begin{equation}\label{eq:reward_n}
% %r[n] = -\Omega + \alpha \cdot \Delta \bar{G} + \beta \cdot \Delta G_{p}
% r[n] = -\Omega + \theta  \times  \mathrm{Gain}
% \end{equation}
% \begin{equation}\label{eq:reward_n}
% %r[n] = -\Omega + \alpha \cdot \Delta \bar{G} + \beta \cdot \Delta G_{p}
% \mathrm{Gain} =  \alpha \cdot \Delta \bar{G} + \beta \cdot \Delta G_{p}
% \end{equation}
% \begin{equation}\label{eq:gain_average}
% \Delta \bar{G} = \frac{\bar{G}[n] - \bar{G}[n-1]}{\bar{G}[n-1]}
% \end{equation}
% \begin{equation}\label{eq:gain_percentile}
% \Delta G_{p} = \frac{G_{p}[n] - G_{p}[n-1]}{G_{p}[n-1]}
% \end{equation}

%$r[n] = -\Omega + \alpha \cdot \Delta \bar{G} + \beta \cdot \Delta G_{p25}$, where $\Delta \bar{G} = \frac{\bar{G}[n] - \bar{G}[n-1]}{\bar{G}[n-1}$ represents the relative average throughput gain since the last  timestep,  $\Delta G_{p25} = \frac{G_{p25}[n] - G_{p25}[n-1]}{G_{p25}[n-1]}$ represents the relative gain of the throughput´s percentile 25 since the last  timestep, 

\noindent where $\Delta G_{k}$ represents the relative gain of the reward factor $k$ in the cell, $\mathbf{B} = \left( B_1,..,B_K \right)$ is the vector of weights with length $K$ and $\theta$ is a scale factor meant to expand the value range of the reward to help with training. Additionally, to help stabilize the training, the reward peak values are limited by the factor $r_\mathrm{clip}$ as follows:

\begin{equation}\label{eq:clip}
r[n] = \mathrm{max}(\mathrm{min}( -|r_\mathrm{clip}|, r[n]), \lvert r_\mathrm{clip}\rvert).
\end{equation}

\section{Simulation results} 
\label{Simulationresults}

%\hl{VAMOS A MOSTRAR}:
%\begin{itemize}
%    \item  - CDF del Throughput en la %evaluacion con SC, CP y el final
%     \item - TABLA con los umbrales e histeresis finales
% \item - GANANCIA de la solución sobre las dos %formas de onda
 %\item - Igual que el anterior pero en tabla
 %\item - Evolución del %reward/histeresis/theshold (???)
 
%\end{itemize}

The environment has been simulated using the 5G toolbox of MATLAB, whereas the DRL agent has been implemented in Keras \& Tensorflow. 
%The DRL agent was trained with 43 different episodes with 75 steps per episode, giving a total of 3225 steps. At each episode, $50$ UEs are randomly dropped around an omnidirectional cell at the origin of the coordinate system. Within each episode, the set of UEs remain unchanged. At each step $n$ all UEs perform an UL transmission that last $1000$ slots. 
The detailed parameters of the DRL training are summarized in Table \ref{table:RLParamsTable}. 
During the length of the transmission, the UEs are bounded to the DPWS following the procedure described in \textbf{Algorithm \ref{alg:DPWSalg}} according to the values of $\zeta$ and $\xi$  of the step. UEs are also subjected to adaptive modulation and coding (AMC), therefore the employed MCS is susceptible to change as the SNR evolves. At any given slot, UEs reporting SNR values below the threshold of the lowest MCS will fall into outage. UEs resulting in outage for the complete duration of the transmission were not scheduled thus not considered in the derived metrics.
%Within an episode, the UEs do not change across steps and their channel realizations are repeated. Said condition implies that observed changes in the throughput values can then only be induced by tweaking the parameters $\zeta$ and $\xi$ via the RL agent. 
%This allows to obtain the statistics and metrics that form the state and reward, which are used by the agent in the next step. 

  During CP-OFDM tranmissions, the precoding matrix $\textbf{W}$ is selected by the BS from the corresponding codebook according to the SRS signal received in order to maximize the received SNR. RL-related parameters can be found in Table \ref{table:RLParamsTable}. All UEs share the same configuration parameters, which are summarized in Table \ref{table:SIMParamsTable}.

\begin{table}[h!]
%\scriptsize
\centering
\caption{RL parameters setting}
\label{table:RLParamsTable}
\begin{tabular}{|l|l|}
\noalign{\hrule height 1pt}
\textbf{Parameter} & \textbf{Value}\\ 
\noalign{\hrule height 1pt}
 $[\epsilon_0, \epsilon_\mathrm{min}]$  & [1, 0.01]  \\
 Learning rate & 0.05\\
Training steps per episode & 75\\
\# of training episodes & $43$\\
Evaluation steps per episode & 20\\
\# of evaluation episodes & $16$\\
UEs per episode & 50\\
$n_\mathrm{hidden}$ & 60 \\
 Optimizer &	Adam \\
 Discount factor & 0.01\\
 Batch size & 350 \\
 Experience buffer size & 750\\
Default threshold, $\zeta$ (dB) & 0\\
Default hysteresis, $\xi$ (dB) & 5\\  
$(\theta, r_\mathrm{clip})$ & $(50, 2)$ \\
\noalign{\hrule height 0.5pt}
\end{tabular}
\end{table}
 % ------------------

\begin{table}[h!]
%\scriptsize
\centering
\caption{Network parameters setting}
\label{table:SIMParamsTable}
\begin{tabular}{|l|l|}
\noalign{\hrule height 1pt}
\textbf{Parameter} & \textbf{Value}\\ 
\noalign{\hrule height 1pt}
 $N$ (RBs) & $20$    \\  
Transmission length (slots) & $1000$    \\  
%Numerology (\mu) & 0 \\
Carrier frequency (GHz)& $28$\\
Subcarrier Spacing (kHz)& $15$\\
User speed  (km/h) & $0.4$ (stationary) \\ 
Delay Spread (ns) & $30$ \\
Channel Delay Profile & CDL-A\\
Pathloss model & Urban Macro\\  
SRS Periodicity (ms) & $2$\\  
Min/max UE to BS distance (m) & $25/300$\\  
$(A_\mathrm{sat}, p)$ & $(24$ dBm, $2)$ \\
$P_\mathrm{max}$ & $23$ dBm \\
\noalign{\hrule height 0.5pt}
\end{tabular}
\end{table}

% [FRAN] SEGUIR!!

To accurately capture the performance in the edge cell, the reward includes $K$ reward factors, $G_k$, associated with low throughput percentiles. 
More specifically, we have considered $K=8$ reward factors, where the reward factor $G_1$ is associated with the $10\%$ percentile, $G_2$ with $15\%$ percentile, and so on up to $G_8$ with $45\%$ percentile (see Table \ref{table:reward_factors}). 
Some bigger percentiles closer to the median are also included as reward factors to avoid harming the average throughput excessively. It was observed during training that relative gains in the lower percentiles were disproportionately bigger compared to those closer to the median. The use of bigger weights for bigger percentiles in the reward function prevents the agent from excessively favoring the lower percentiles and damaging the average cell throughput in the process. Table \ref{table:reward_factors} summarizes all selected reward factors $G_k$ and their associated weights $B_k$.

% \begin{table}[h!]
% %\scriptsize
% \centering
% \caption{Reward factors and their corresponding weights}
% \label{table:RLParamsTable}
% \begin{tabular}{|l|l|}
% \noalign{\hrule height 1pt}
% \textbf{$G_k$} & \textbf{$W_k$}\\ 
% \noalign{\hrule height 1pt}
% p10	&	0.02	\\
% p15	&	0.04		\\
% p20	&	0.06	\\
% p25	&	0.08		\\
% p30	&	0.10		\\
% p35	&	0.12		\\
% p40	&	0.14		\\
% p45	&	0.16		\\
% Avg	&	0.18		\\
% \noalign{\hrule height 0.5pt}

\begin{table}[h!]
\centering
\caption{Reward factors and their corresponding weights}
\label{table:reward_factors}
\begin{tabular}{|c|c|c|c|c|c|c|c|c|c|}
\hline
\textbf{$G_k$} & p10 & p15 & p20 & p25 & p30 & p35 & p40 & p45 & Avg \\
\hline
\textbf{$B_k$} & 0.02 & 0.04 & 0.06 & 0.08 & 0.10 & 0.12 & 0.14 & 0.16 & 0.18 \\
\hline
\end{tabular}
\end{table}

After the training stage was completed, the agent was evaluated on 16 different episodes. Given the incremental nature of the action space, the agent was given $20$ steps to reach the solution for each episode. It is worth noting that while the UEs in the evaluation stage are drawn from the same configuration as the training stage, the drops, and therefore their positions and channel realizations, are completely independent.

Table \ref{table:Results_table} summarizes the achieved throughput gains of the AI-DPWS waveform switching in contrast to fixed waveform schemes. Throughput gains are expressed in relative and absolute terms for all reward factors $G_k$. These results are drawn from the cumulative metrics of all UEs in the 16 evaluation episodes. From the presented table it is easy to observe that the proposed scheme is able to achieve higher throughput for almost all the different reward factors across both fixed waveform schemes. It is worth noting that the bigger relative performance gains are observed towards the lower percentiles even if the absolute gains of all reward factors are comparable. Interestingly, these gains in the lower percentiles do not come as a performance loss in the average throughput of the cell.

\begin{table}[h!]
%\scriptsize
\centering
\caption{Evaluation results}
\label{table:Results_table}
\begin{tabular}{|c|c|c|c|c|}
\noalign{\hrule height 0.5pt}
% & \multicolumn{2}{|c|}{Throughput (Mbps)} & \multicolumn{2}{|c|}{Throughput (Mbps)} & \multicolumn{2}{|c|}{Throughput} & \\
& \multicolumn{4}{|c|}{Mean throughput gain over:}  \\
\noalign{\hrule height 0.5pt}
& \multicolumn{2}{|c|}{CP-OFDM} & \multicolumn{2}{|c|}{DFT-S-OFDM}  \\
\noalign{\hrule height 0.5pt} \\[-0.9em]
\textbf{$G_k$} & \% & Abs (Mbps) & \% & Abs (Mbps)\\

\noalign{\hrule height 0.5pt} \\[-0.9em]

p10	&	297.9299	&	0.0049	&	216.8994	&	0.0046	\\
p15	&	42.393	&	0.0159	&	-16.6991	&	-0.0125	\\
p20	&	27.2969	&	0.0129	&	7.5623	&	0.0086	\\
p25	&	7.9753	&	0.0122	&	10.9392	&	0.0173	\\
p30	&	8.7799	&	0.0129	&	2.0228	&	0.0046	\\
p35	&	3.6767	&	0.0048	&	3.1608	&	0.0198	\\
p40	&	2.3618	&	0.0067	&	4.9113	&	0.0311	\\
p45	&	-0.2176	&	-0.0028	&	6.4336	&	0.0584	\\
Avg	&	0.116	&	0.002	&	3.3794	&	0.0603	\\
\noalign{\hrule height 0.5pt}

\end{tabular}
 \label{table:gains}
\end{table}

\begin{figure}[t!]%
    \centering
    \subfloat[]{\includegraphics[width=0.6\columnwidth]{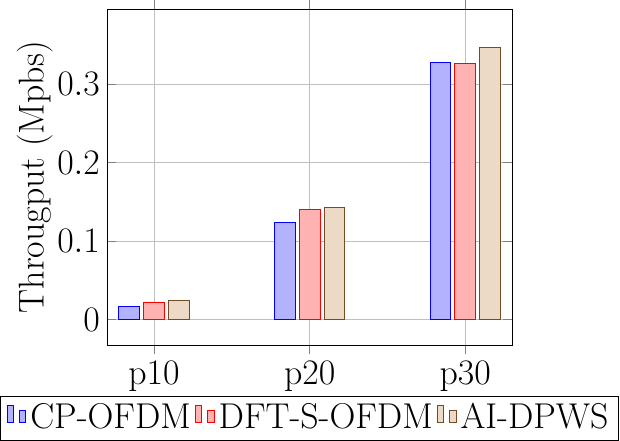} }%
    \subfloat[]{\includegraphics[width=0.45\columnwidth]{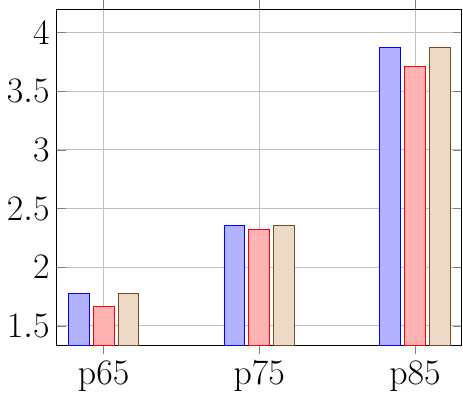} }%
    \caption{Achieved throughput with both fixed waveforms and with AI-assisted switching for: (a) low percentiles; and (b) high percentiles.}%
    \label{fig:percentiles}%
\end{figure}

Finally, to better visualize the impact of AI-DPWS across all UEs in the cell, Fig. \ref{fig:percentiles} (a) and (b) illustrate the achieved throughput of both fixed waveforms and AI-DPWS for low and high throughput percentiles respectively. 

As expected, for the lower percentiles, i.e., the cell-edge UEs, DFT-S-OFDM outperforms CP-OFDM. However, AI-DPWS outperforms DFT-S-OFDM by taking into account the conditions of the channel and switching the waveforms accordingly. For higher percentiles, i.e. cell-interior UEs, CP-OFDM outperforms DFT-S-OFDM and in this case, AI-DPWS can only match the performance of the best fixed waveform scheme.

\section{Conclusions} 
\label{conclusions}

In this paper, a DRL-based framework is proposed to dynamically select the optimal threshold and hysteresis for UL waveform selection based on the DPWS scheme. By using a realistic 5G simulator and realistic measurements available in current networks, it is shown that the proposed scheme outperforms fixed schemes where the waveforms are selected in a static manner. 
Such performance improvements are exhibited across a wide range of throughput percentiles, showing bigger gains in the lower percentiles, which accounts for the cell-edge UEs. These improvements come without any harm to the average or even top-performing UEs in the cell.

\bibliographystyle{IEEEtran}
\bibliography{references.bib}

\begin{thebibliography}{10}
\providecommand{\url}[1]{#1}
\csname url@samestyle\endcsname
\providecommand{\newblock}{\relax}
\providecommand{\bibinfo}[2]{#2}
\providecommand{\BIBentrySTDinterwordspacing}{\spaceskip=0pt\relax}
\providecommand{\BIBentryALTinterwordstretchfactor}{4}
\providecommand{\BIBentryALTinterwordspacing}{\spaceskip=\fontdimen2\font plus
\BIBentryALTinterwordstretchfactor\fontdimen3\font minus
  \fontdimen4\font\relax}
\providecommand{\BIBforeignlanguage}[2]{{%
\expandafter\ifx\csname l@#1\endcsname\relax
\typeout{** WARNING: IEEEtran.bst: No hyphenation pattern has been}%
\typeout{** loaded for the language `#1'. Using the pattern for}%
\typeout{** the default language instead.}%
\else
\language=\csname l@#1\endcsname
\fi
#2}}
\providecommand{\BIBdecl}{\relax}
\BIBdecl

\bibitem{ref:MPR}
3GPP, \emph{{User Equipment (UE) radio transmission and reception; Part 2:
  Range 2 Standalone (Release 17)}}, {3rd Generation Partnership Project
  (3GPP)} TS {38.101-2}, Rev. 17.9.0, April 2023.

\bibitem{ref:Zheng09}
K.~Zheng and et~al., ``Impacts of amplifier nonlinearities on uplink
  performance in 3g lte systems,'' in \emph{2009 Fourth International
  Conference on Communications and Networking in China}, 2009.

\bibitem{ref:WO2021/047973}
R.~Ramirez-Gutierrez and A.~Nader, ``Switching waveforms for uplink
  transmission in nr network,'' Mar 2021, patent number US2022376965A1.

\bibitem{ref:Feriani21}
A.~Feriani and E.~Hossain, ``Single and multi-agent deep reinforcement learning
  for ai-enabled wireless networks: A tutorial,'' \emph{IEEE Communications
  Surveys \& Tutorials}, vol.~23, no.~2, pp. 1226--1252, 2021.

\bibitem{Ahsan21}
W.~Ahsan and et~al., ``Resource allocation in uplink noma-iot networks: A
  reinforcement-learning approach,'' \emph{IEEE Trans. on Wireless Comm.},
  vol.~20, no.~8, 2021.

\bibitem{Neto21}
N.~Costa and et~al., ``Uplink power control framework based on reinforcement
  learning for 5g networks,'' \emph{IEEE Trans. on Vehicular Technology},
  vol.~70, no.~6, 2021.

\bibitem{Teng19}
Y.~Teng and et~al., ``Distributed learning solution for uplink traffic control
  in energy harvesting massive machine-type communications,'' \emph{IEEE
  Wireless Communications Letters}, vol.~9, no.~4, 2020.

\bibitem{ref:Zaidi}
A.~A. Zaidi and et~al., ``A preliminary study on waveform candidates for {5G}
  mobile radio communications above 6 {GHz},'' in \emph{2016 IEEE 83rd
  Vehicular Technology Conference (VTC Spring)}, 2016.

\bibitem{ref:RappAmp}
C.~{Rapp}, ``{Effects of HPA-nonlinearity on a 4-DPSK/OFDM-signal for a digital
  sound broadcasting signal},'' in \emph{ESA Special Publication}, ser. ESA
  Special Publication, B.~{Kaldeich}, Ed., vol. 332, Oct. 1991, pp. 179--184.

\bibitem{ref:38.331}
3GPP, \emph{{Technical Specification Group Radio Access Network; NR; Radio
  Resource Control (RRC) protocol specification (Release 17)}}, {3rd Generation
  Partnership Project (3GPP)} TS {38.331}, Rev. 17.4.0, May 2023.

\end{thebibliography}

\end{document}